\documentstyle[prb,aps,psfig,twocolumn]{revtex}

\begin{document}
\twocolumn[\hsize\textwidth\columnwidth\hsize\csname
@twocolumnfalse\endcsname

\title
{
Charge and orbital ordering 
in underdoped La$_{1-x}$Sr$_x$MnO$_3$
}
\author{T. Mizokawa$^{1,2}$, D. I. Khomskii$^{1}$, 
and G. A. Sawatzky$^{1}$}
\address{
$^{1}$Solid State Physics Laboratory,
Materials Science Centre,
University of Groningen,
Nijenborgh 4, 9747 AG Groningen,
The Netherlands}
\address{
$^{2}$Department of Complexity Science and Engineering,
University of Tokyo, Bunkyo-ku, Tokyo 113-0033, Japan
}
\date{\today}
\maketitle

\begin{abstract}
We have explored spin, charge and orbitally ordered states
in La$_{1-x}$Sr$_x$MnO$_3$ ($0 < x < 1/2$) 
using model Hartree-Fock calculations on $d$-$p$-type lattice models.
At $x$=1/8, several charge and orbitally modulated states are found
to be stable and almost degenerate in energy with a homogeneous
ferromagnetic state. The present calculation indicates
that a ferromagnetic state with a charge modulation along the $c$-axis
which is consistent with the experiment by Yamada {\it et al.} 
might be responsible for the anomalous behavior around $x$ = 1/8.
\end{abstract}
]

La$_{1-x}$Sr$_x$MnO$_3$ have extensively been studied 
because of its interesting magnetic and electric properties
\cite{Goodenough,Tokura}. An antiferromagnetic insulator 
LaMnO$_3$ evolves into a ferromagnetic metal 
with substitution of Sr for La or with hole doping \cite{Tokura}.
Underdoped La$_{1-x}$Sr$_x$MnO$_3$ with $x$ $\sim$ 1/8 
which is located between the antiferromagnetic
insulating region and the ferromagnetic metallic region,
shows many anomalous behaviors
\cite{Tokura,Kawano,Yamada,Loidl,Endoh}. 
La$_{0.875}$Sr$_{0.125}$MnO$_3$ 
is a paramagnetic insulator above $T_{CA}$ (180 K) 
and has a canted antferromagnetic state below it \cite{Kawano}. 
Recently, it has been found that La$_{0.875}$Sr$_{0.125}$MnO$_3$
becomes a ferromagnetic insulator below 140 K \cite{Loidl,Endoh}.
One important question is why the hole-doped system can exist as
an insulator. The superstructure observed by Yamada {\it et al.}
\cite{Yamada} indicates that charge ordering 
is responsible for the insulating behavior. 
However, it is still controversial whether charge ordering is
realized in La$_{0.875}$Sr$_{0.125}$MnO$_3$ or not.
Ahn and Millis studied the charge and orbital ordering in
La$_{0.875}$Sr$_{0.125}$MnO$_3$ using a model of 
strong electron-lattice coupling limit and 
found that the charge ordering proposed by Yamada {\it et al.} 
can be reproduced using their model \cite{Millis}.
On the other hand, using the resonant x-ray scattering,
Endoh {\it et al.} confirmed that the superlattice peak 
found by Yamada {\it et al.} does not show resonance 
at the Mn $K$-edge and concluded that
there is no Mn$^{3+}$/Mn$^{4+}$ charge ordering 
in La$_{0.875}$Sr$_{0.125}$MnO$_3$ \cite{Endoh}.
Another interesting question is what is the origin
of the ferromagnetism. Since the system is insulating,
the simple double exchange mechanism cannot be applied.
A model Hartree-Fock (HF) calculation for LaMnO$_3$ has predicted 
that, if the Jahn-Teller distortion is suppressed, 
a ferromagnetic insulating state with orbital ordering 
would be realized \cite{HF}. This means that the superexchange 
interaction between the Mn$^{3+}$ ions can be ferromagnetic 
because of orbital ordering \cite{KK}. 
Endoh {\it et al.} observed orbital ordering below 145 K
using x-ray scattering technique and argued that 
the orbital ordering is essential for the ferromagnetic 
and insulating state \cite{Endoh}. 
However, the orbital ordered state without charge ordering
is expected to be metallic and may not be consistent with the fact
that La$_{0.875}$Sr$_{0.125}$MnO$_3$ is insulating.
In the doped manganites, the orbital modulation
should couple with the charge modulation in a similar way that
the spin modulation couples with the charge modulation
in the doped cuprates \cite{Koshibae}. 
In order to understand the electronic
structure of the ferromagnetic and insulating state 
in the doped manganites, it is necessary to consider 
the complicated interplay between the charge and orbital orderings.
In this paper, we study possibility of charge 
and orbitally ordered states in underdoped La$_{1-x}$Sr$_x$MnO$_3$
using the model HF calculation and explore 
the origin of the ferromagnetic insulating state
in underdoped La$_{1-x}$Sr$_x$MnO$_3$.

We use the multi-band $d$-$p$ model with 16 Mn 
and 48 oxygen sites in which full degeneracy of Mn 3$d$ orbitals
and the oxygen 2$p$ orbitals are taken into account
\cite{HF}.
The Hamiltonian is given by\\
\begin{eqnarray}
H = H_p + H_d + H_{pd},
\end{eqnarray}
\begin{eqnarray}
H_p = {\displaystyle \sum_{k,l,\sigma}}
\epsilon^p_{k} p^+_{k,l\sigma}p_{k,l\sigma}
+ {\displaystyle \sum_{k,l>l',\sigma} 
V^{pp}_{k,ll'} p^+_{k,l\sigma}p_{k,l'\sigma}}
+ H.c.,
\end{eqnarray}
\begin{eqnarray}
H_d & = & \epsilon_d {\displaystyle \sum_{i,m\sigma}}
d^+_{i,m\sigma}d_{i,m\sigma}
+ u {\displaystyle \sum_{i,m}}
d^+_{i,m\uparrow}d_{i,m\uparrow}d^+_{i,m\downarrow}d_{i,m\downarrow}
\nonumber \\
& + & u' {\displaystyle \sum_{i,m \neq m'}}
d^+_{i,m\uparrow}d_{i,m\uparrow}d^+_{i,m'\downarrow}d_{i,m'\downarrow}
\nonumber \\
& + & (u'-j') {\displaystyle \sum_{i,m>m',\sigma}}
d^+_{i,m\sigma}d_{i,m\sigma}d^+_{i,m'\sigma}d_{i,m'\sigma}
\nonumber \\
& + & j' {\displaystyle \sum_{i,m \neq m'}}
d^+_{i,m\uparrow}d_{i,m'\uparrow}d^+_{i,m\downarrow}d_{i,m'\downarrow}
\nonumber \\
& + & j {\displaystyle \sum_{i,m \neq m'}}
d^+_{i,m\uparrow}d_{i,m'\uparrow}d^+_{i,m'\downarrow}d_{i,m\downarrow},
\end{eqnarray}
\begin{eqnarray}
H_{pd} = {\displaystyle \sum_{k,m,l,\sigma}} V^{pd}_{k,lm}
d^+_{k,m\sigma}p_{k,l\sigma} + H.c.
\end{eqnarray}
$d^+_{i,m\sigma}$ are creation operators for the 3$d$ electrons
at site $i$.
$d^+_{k,m\sigma}$ and $p^+_{k,l\sigma}$ are creation operators
for Bloch electrons with wave vector $k$ 
which are constructed from the $m$-th component 
of the 3$d$ orbitals and from the $l$-th component 
of the 2$p$ orbitals, respectively.
The intra-atomic Coulomb interaction between the 3$d$ electrons
is expressed using
Kanamori parameters, $u$, $u'$, $j$ and $j'$ \cite{Kanamori}.
The transfer integrals between Mn 3$d$ 
and oxygen 2$p$ orbitals $V^{pd}_{k,lm}$ 
are given in terms of Slater-Koster parameters
$(pd\sigma)$ and $(pd\pi)$. The transfer integrals
between the oxygen 2$p$ orbitals
$V^{pp}_{k,ll'}$ are expressed by $(pp\sigma)$ and $(pp\pi)$.
Here, the ratio $(pd\sigma)$/$(pd\pi)$ is -2.16.
$(pp\sigma)$ and $(pp\pi)$ are fixed at -0.60 and 0.15,
respectively, for the undistorted lattice.
When the lattice is distorted, the transfer integrals 
are scaled using Harrison's law \cite{Harrison}.
The charge-transfer energy $\Delta$ is defined by
$\epsilon^0_d - \epsilon_p + nU$, where
$\epsilon^0_d$ and $\epsilon_p$ are the energies of the bare
3$d$ and 2$p$ orbitals and $U$ ($=u -20/9j$) 
is the multiplet-averaged $d-d$ Coulomb interaction.
$\Delta$, $U$, and $(pd\sigma)$ for LaMnO$_3$ 
are 4.0, 5.5, and -1.8 eV, respectively, which are
taken from the photoemission study \cite{Saitoh}. 

\begin{figure}
\psfig{figure=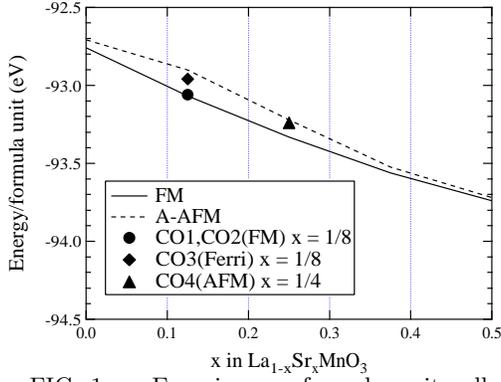,width=7cm}
\caption{
Energies per formula unit cell
of the various charge-ordered states, 
the ferromagnetic state (solid curve), and
the $A$-type antiferromagnetic state (dashed curve) 
as functions of hole concentration $x$.
} 
\label{TE}
\end{figure}

\begin{figure}
\psfig{figure=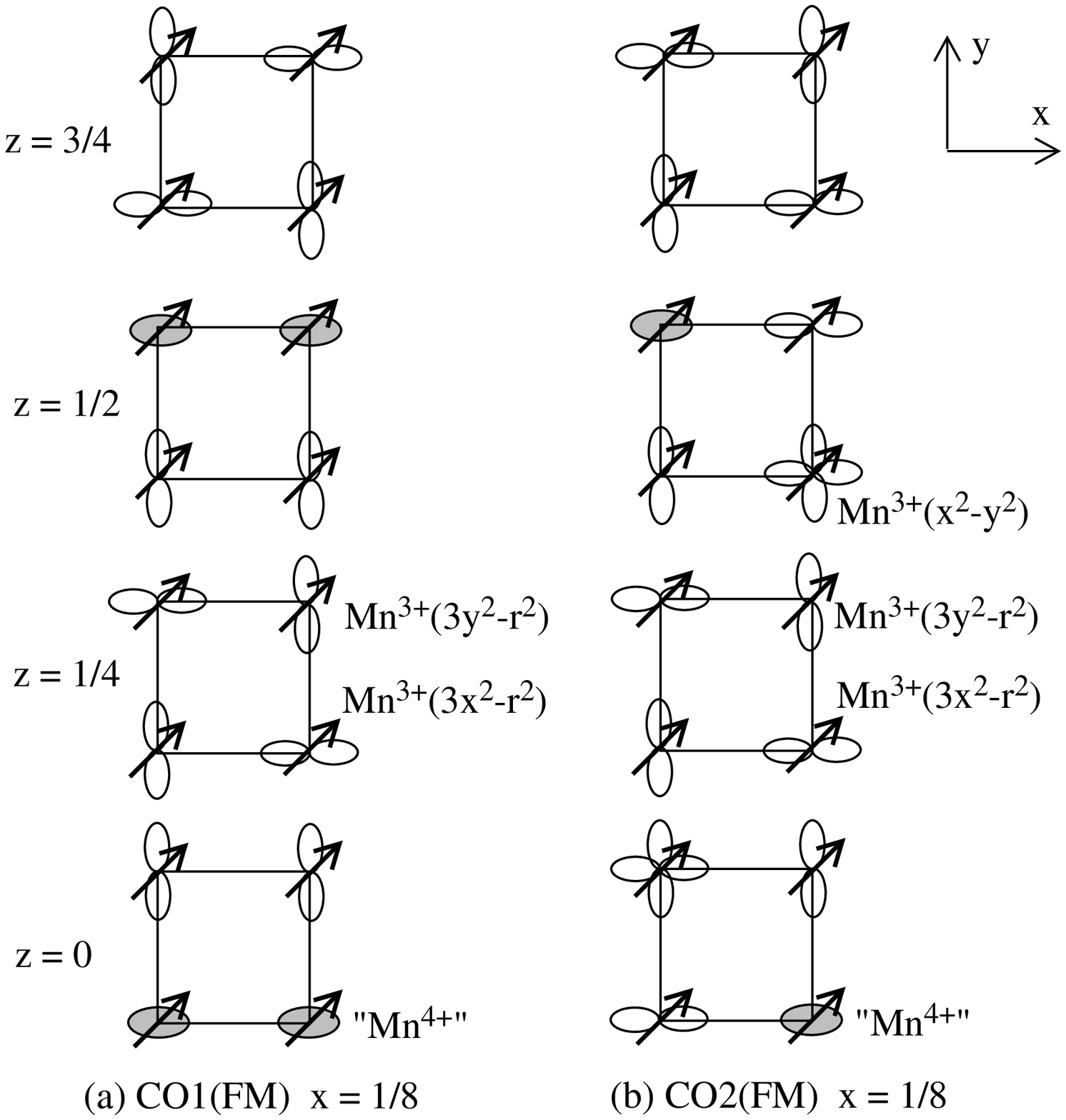,width=7cm}
\caption{
Schematic drawings of the charge and orbital orderings 
for the ferromagnetic state at $x$ =1/8.
The unit cell consists of four layers of $z$ =0, 1/4, 1/2,
and 3/4, each of which have four Mn sites.
The arrows indicate spin directions at each site and 
the shaded orbitals are for Mn$^{4+}$-like hole-rich sites.
}
\label{CO}
\end{figure}

In Fig. \ref{TE}, the energies of the spin, charge and orbitally 
ordered states are compared with those of 
the ferromagnetic and $A$-type antiferromagnetic states, 
which are plotted as functions of the hole concentration $x$.
At $x$ of 1/8, several charge ordered states 
exist as stable solutions.
A schematic drawings of the ferromagnetic charge-ordered states 
are shown in Fig. \ref{CO}. The unit cell consists of the
four layers of $z$ = 0, 1/4, 1/2, and 3/4 along the $c$-axis. 
Each layer has four different Mn sites.
In the charge-ordered states, 
the hole-rich planes ($z$ = 0 and 1/2) and 
the hole-poor planes ($z$ = 1/4 and 3/4) are 
alternatingly stacked along the $c$-axis.
In the hole-poor plane, either $d_{3x^2-r^2}$-like or
$d_{3y^2-r^2}$-like orbital is mainly occupied at each site
and the $3x^2-r^2/3y^2-r^2$-type orbital ordering.
This orbital ordering in the hole-poor plane 
is essentially the same as that in LaMnO$_3$ 
although it is weak compared to that in LaMnO$_3$. 
While the orbital orderings at $z$ = 0 and at $z$ = 3/4
are out of phase for CO1 [see Fig. \ref{CO}(a)], 
those are in phase for CO2 as shown in Fig. \ref{CO}(b).
Probably, the orbital ordering in the hole-poor plane 
is related to the observation by Endoh {\it et al.} \cite{Endoh}.
In the hole-rich plane, for CO1, the Mn$^{4+}$-like sites form 
a kind of stripe as shown in Fig. \ref{CO}(a) and the extra holes
are sitting at the oxygen sites between the Mn$^{4+}$-like sites.
For the CO2 state, the Mn$^{4+}$-like sites form a square lattice
and the extra holes are distributed at the oxygen sites 
surrounding the Mn$^{4+}$-like site.
The orbital ordering at the Mn$^{3+}$-like sites 
in the hole-rich plane is very weak and depends on 
the orbital ordering in the hole-poor plane.
These two ferromagnetic states with the charge 
and orbital modulations are degenerate in energy 
within the accuracy of the present calculation 
and are the lowest in energy among the charge-ordered states
obtained in the present model calculations.
Since LaMnO$_3$ is a charge-transfer-type Mott insulator, 
the Mn$^{4+}$-like site has approximately four electrons.
For example, in the CO1 state,
The number of $3d$ electrons at the Mn$^{3+}$-like sites 
in the hole-poor plane is $\sim$ 4.08 and that of the Mn$^{3+}$-like 
and the Mn$^{4+}$-like sites in the hole-rich plane 
are $\sim$ 4.03 and $\sim$ 4.01, respectively. 
This calculated result can explain why
the resonant x-ray scattering cannot distinguish
between the Mn$^{3+}$-like and Mn$^{4+}$-like sites \cite{Endoh}.

\begin{figure}
\psfig{figure=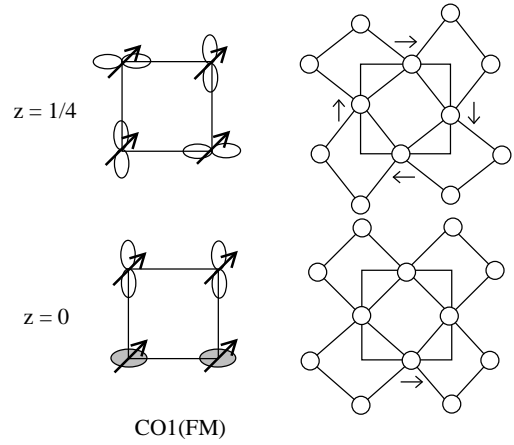,width=7cm}
\caption{
Schematic drawings of the charge and orbital
orderings (left column) and the lattice distortions 
(right column) for the hole-rich ($z$ = 0) and hole-poor
($z$ = 1/4) planes in the ferromagnetic state.
The open circles and the arrows indicate 
the oxygen ions and the shifts of the oxygen ions, respectively.
}
\label{DIS}
\end{figure}

These ferromagnetic states with the charge modulations
are still metallic without lattice distortion. 
We have studied the effect of the lattice distortion
which is shown in the right column of Fig. \ref{DIS}.
The hole-poor plane ($z$ = 1/4) can couple with 
the Jahn-Teller distortion of LaMnO$_3$. On the other hand,
in the hole-rich plane ($z$ = 0) for CO1, the shift of oxygen 
ion sitting between the Mn$^{4+}$-like sites causes 
a doubling along the stripe of the Mn$^{4+}$-like sites 
and is expected to open a band gap.
Actually, we found that the small shift of these oxygens 
less than 0.1 $\AA$ (Fig. \ref{DIS}), 
which gives the superstructure along the $c$-axis 
and is consistent with the experiment 
by Yamada {\it et al.} \cite{Yamada}, can open a band gap 
for the two charge-modulated ferromagnetic states.
The lattice distortion shwon in Fig. \ref{DIS} is enough
to open a band gap for the CO2 state although
the breathing-type distortion at the Mn$^{4+}$-like sites is
expected to be more effective.
These ferromagnetic states with the lattice distortions
are strong candidates for the ferromagnetic and insulating state 
found in La$_{0.875}$Sr$_{0.125}$MnO$_3$.
 
\begin{figure}
\psfig{figure=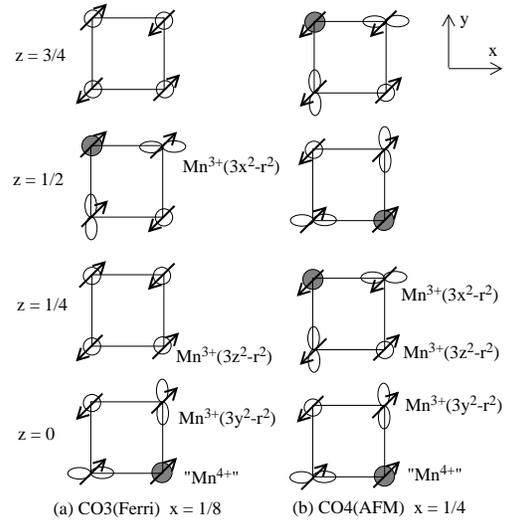,width=7cm}
\caption{
Schematic drawings of the charge and orbital orderings 
(a) for the ferrimagnetic states at $x$ =1/8 and
(b) for the antiferromagnetic state at $x$ = 1/4.
The unit cell consists of four layers of $z$ =0, 1/4, 1/2,
and 3/4, each of which have four Mn sites.
The arrows indicate spin directions at each site and 
the shaded orbitals are for Mn$^{4+}$-like hole-rich sites.
}
\label{CO2}
\end{figure}

\begin{figure}
\psfig{figure=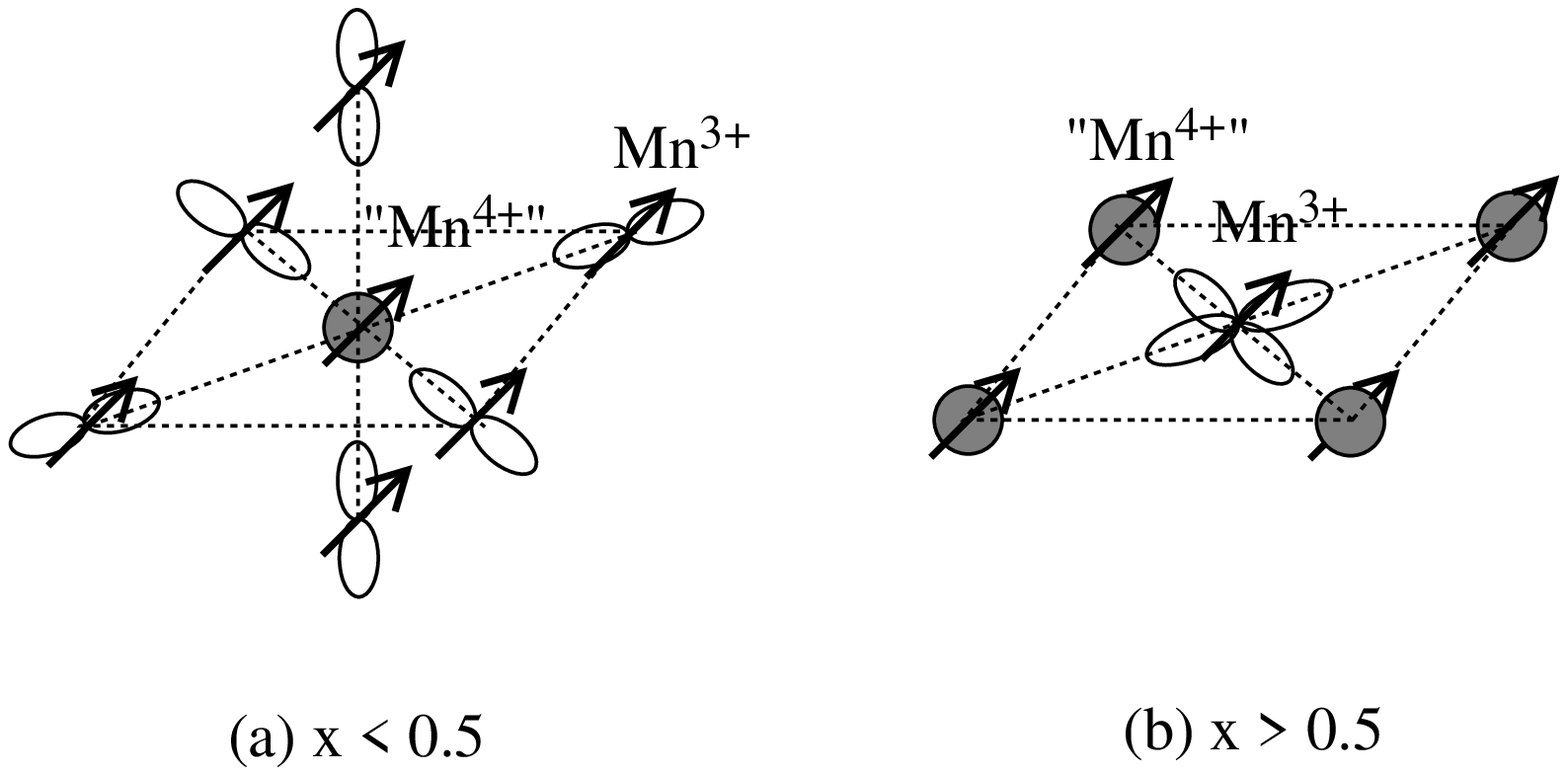,width=7cm}
\caption{
Schematic drawings of the orbital polarons
(a) for $x < 0.5$ and (b) for $x > 0.5$.
The arrows indicate spin directions at each site and 
the shaded orbitals are for Mn$^{4+}$-like hole-rich sites.
}
\label{OP}
\end{figure}

A sketch of the ferrimagnetic state accompanied by the charge
and orbital ordering is shown in Fig. \ref{CO2}(a).
In this arrangement, each hole-rich  Mn$^{4+}$-like site 
is surrounded by six hole-poor Mn$^{3+}$ sites. 
This can be viewed as a lattice of orbital polaron \cite{Loidl2} 
which is displayed in Fig. \ref{OP}(a).
The superexchange interaction between the Mn$^{3+}$-like
and Mn$^{4+}$-like sites is ferromagnetic and that between
the Mn$^{3+}$-like sites is antiferromagnetic 
in this ferrimagnetic state. The orbital polaron 
might be related to the $\sim 12 \AA$ magnetic clusters 
observed in La$_{0.67}$Ca$_{0.33}$MnO$_3$ \cite{MP}.
It is expected that the existence of this orbital polaron 
makes the magnetic interaction along the $c$-axis 
ferromagnetic and gives three-dimensional ferromagnetic coupling.
Since the Mn$^{3+}$-like sites should be accompanied by 
the Jahn-Teller distortion, the orbital polaron might 
also be related to the Jahn-Teller polaron observed 
in La$_{1-x}$Ca$_{x}$MnO$_3$ with $x < 0.5$ \cite{JT}.
On the other hand, in La$_{1-x}$Ca$_{x}$MnO$_3$ with $x > 0.5$,
the number of Mn$^{4+}$-like sites is larger than 
that of Mn$^{3+}$-like sites. In such a case, the orbital
polaron, in which a Mn$^{4+}$-like site is
surrounded by four Mn$^{3+}$-like sites [see Fig. \ref{OP}(b)],
is expected to be relevant.
This orbital polaron gives two-dimensional ferromagnetic coupling.
Actually, it has been reported that Nd$_{1-x}$Sr$_{x}$MnO$_3$ 
with $x > 0.5$ has the $A$-type antiferromagnetic state, 
in which the ferromagnetic $ab$-planes are antiferromagnetically 
coupled along the $c$-axis \cite{Kawano2}. This orbital polaron
might be relevant in the $A$-type antiferromagnetic state
of Nd$_{1-x}$Sr$_{x}$MnO$_3$.

At $x$ = 1/4, the homogeneous ferromagnetic state is 
very stable and no ferromagnetic charge-modulated state was 
obtained. However, it is found that an antiferromagnetic 
state with charge and orbital ordering exists as a stable solution
which is schematically shown in Fig. \ref{CO2}(b).
This charge-ordered state can be viewed as a lattice of the orbital
polaron coupled antiferromagnetically. Although this state is higher 
in energy than the homogeneous ferromagnetic state as shown 
in Fig. \ref{TE}, a lattice distortion of breathing type 
may stabilize the charge-ordered state relative to 
the ferromagnetic state.

There are two possible ways to describe the charge-ordered
states: a polaron lattice language and a charge-density wave language.
When the electron-lattice coupling term dominates the other terms, 
the polaron lattice picture is appropriate.
The ferrimagentic charge-ordered state obtained above can be viewed
as a orbital polaron lattice and is expected to strongly
couple with lattice distortions or the Jahn-Teller distortions.
On the other hand, the charge-density-wave language 
becomes more appropriate when the kinetic energy term is relevant.
It is natural to speculate that the electron-lattice coupling
is weak in the ferromagnetic charge-modulated state 
compared to the orbital polaron lattice. 
The present calculation neglecting lattice distortions 
indicates that a kind of Umklapp process can 
give the modulation along the $c$-axis and that 
the charge-density-wave picture might be relevant
in La$_{0.875}$Sr$_{0.125}$MnO$_3$.
Since the number of 3$d$ electrons at the Mn$^{4+}$-like site
is almost the same as that of the Mn$^{3+}$-like site,
the electron-lattice coupling is expected to be small.
Actually the observed lattice modulation along the $c$-axis
is very small \cite{Yamada}. This is also consistent with the experimental 
observation that resonant x-ray scattering fails to distinguish
between Mn$^{3+}$-like and Mn$^{4+}$-like sites \cite{Endoh}. 
The present calculation fails to give a finite band gap
without extra lattice distortion,
suggesting that the weak electron-lattice coupling is still
important to give the band gap.
Here, it should be noted that a perfect nesting is not required
in the present system because the Coulomb interaction term
between the $3d$ electrons is very large and is comparable 
to the kinetic energy term and that, in this sense, 
the magnetic coupling between two Mn sites can be 
viewed as a kind of superexchange coupling.

In conclusion, we have studied possible charge 
and orbitally ordered states in underdoped La$_{1-x}$Sr$_x$MnO$_3$
using the model HF calculation. It has been found that 
the ferromagnetic state with the charge ordering along the $c$-axis,
which is consistent with the experiment by Yamada {\it et al.}
\cite{Yamada}, is stable at $x$=1/8.
It has been argued that the charge and orbital ordering 
in the ferrimagnetic state can be interpreted 
as orbital polaron lattice.
In order to clarify the interplay between the charge/orbital
ordering and the lattice distortion, the underdoped manganites
should be studied in future using a more realistic model which includes
the electron-lattice interaction.

The authors would like to thank J. L. Garcia-Mu$\tilde{\rm n}$os 
for useful discussions. This work was supported by 
the Nederlands Organization for Fundamental Research 
of Matter (FOM) and by the European Commission TRM network 
on Oxide Spin Electronics (OXSEN).

\begin{references}

\bibitem{Goodenough}
J. B. Goodenough, Phys. Rev. {\bf 100}, 564 (1955);
J. B. Goodenough and J. S. Zou, Nature {386}, 229 (1997).

\bibitem{Tokura}
A. Urushibara, Y. Moritomo, T. Arima, A. Asamitsu,
G. Kido, and Y. Tokura, Phys. Rev. B {\bf 51}, 14103 (1995).

\bibitem{Kawano}
H. Kawano, R. Kajimoto, M. Kubota, and H. Yoshizawa,
Phys. Rev. B {\bf 53}, R14709 (1996).

\bibitem{Yamada}
Y. Yamada, O. Hino, S. Nohdo, R. Kanao, T. Inami, and S. Katano,
Phys. Rev. Lett. {\bf 77}, 904 (1996).

\bibitem{Loidl}
R. Senis, V. Laukhin, B. Mart\'{i}nez, J. Fontcuberta, and X. Obradors, 
Phys. Rev. B {\bf 57}, 14680 (1998);
S. Uhlenbruck, R. Teipen, R. Klingler, B. B\"{u}chner, O. Friedt,
M. H\"{u}cher, H. Kierspel, T. Niem\"{o}ller, L. Pinsard,
A. Revcolevschi, and R. Gross, Phys. Rev. Lett. {\bf 82}, 185 (1999); 
M. Paraskevopoulos, J. Hemberger, A. Loidl, A. A. Mukhin, 
V. Yu Ivanov and A. M. Balbashov, cond-mat/9812276.

\bibitem{Endoh}
Y. Endoh, K. Hirota, S. Ishihara, S. Okamoto, Y. Murakami, 
A. Nishizawa, T. Fukuda, H. Kimura, H. Nojiri, K. Kaneko,
and S. Maekawa, Phys. Rev. Lett. {\bf 82}, 4328 (1999).

\bibitem{Millis}
K. H. Ahn and A. J. Millis, Phys. Rev. B {\bf 58}, 3697 (1998).

\bibitem{HF}
T. Mizokawa and A. Fujimori, Phys. Rev. B
{\bf 51}, 12880 (1995); Phys. Rev. B {\bf 54}, 5368 (1996).

\bibitem{KK}
K. I. Kugel and D. I. Khomskii,
JETP Lett. {\bf 15}, 446 (1972);
Usp. Fiz. Nauk. {\bf 136}, 621 (1981) [Sov. Phys. Usp.
{\bf 25}, 231 (1982).]

\bibitem{Koshibae}
W. Koshibae, Y. Kawamura, S. Ishihara, S. Okamoto,
J. Inoue, and S. Maekawa, J. Phys. Soc. Jpn {\bf 66}, 957 (1997);
V. I. Anisimov, I. S. Elfimov, M. A. Korotin, 
and K. Terakura, Phys. Rev. B {\bf 55}, 15494 (1997);
T. Mizokawa and A. Fujimori, Phys. Rev. B {\bf 56}, R493 (1997).

\bibitem{Kanamori}
J. Kanamori, Prog. Theor. Phys. {\bf 30}, 275 (1963).

\bibitem{Harrison}
W. A. Harrison, {\em Electronic structure
and the Properties of Solids} (Dover, New York, 1989).

\bibitem{Saitoh}
T. Saitoh, A. E. Bocquet, T. Mizokawa, H. Namatame, A. Fujimori, 
M. Abbate, Y. Takeda, and M. Takano,
Phys. Rev. B {\bf 51}, 13942 (1995).

\bibitem{Loidl2}
M. Paraskevopoulos, J. Hemberger, A. Loidl,
A. A. Mukhin, V. Yu. Ivanov, and A. M. Balbashov, 
cond-mat/9812305;
R. Kilian and G. Khaliullin, Phys. Rev. B {\bf 60}, 13458 (1999).

\bibitem{MP}
J. M. De Teresa, M. R. Ibarra, P. A. Algarabel, C. Ritter,
C. Marquina, J. Blasco, J. Garcia, A. del Moral, and Z. Arnold,
Nature {\bf 386}, 256 (1997).

\bibitem{JT}A. Lanzara, N. L. Saini, M. Brunelli,
F. Natali, A. Bianconi, P. G. Radaelli, and S.-W. Cheong,
Phys. Rev. Lett. {\bf 81}, 878 (1998).

\bibitem{Kawano2}
H. Kawano, R. Kajimoto, H. Yoshizawa, Y. Tomioka,
H. Kuwahara, and Y. Tokura,
Phys. Rev. Lett. {\bf 78}, 4253 (1997).

\end{references}
\end{document}